# Native Skywatchers and the *Ojibwe Giizhig Anung Masinaaigan* – Ojibwe Sky Star Map

Annette S. Lee

*Assistant Professor of Astronomy & Physics, 324 Wick Science Building, St. Cloud State University, 720 Fourth Avenue South, St. Cloud, Minnesota 56301*

**Abstract**. The *Native Skywatchers* programming addresses the loss of Ojibwe and Dakota star knowledge. The goal is to help preserve indigenous astronomy and pass it on to future generations. The star knowledge will serve as a cornerstone inspiration for native communities and at the same time be influential to Native students interested in science and math. Presented here is the *Ojibwe Giizhig Anung Masinaaigan* – Ojibwe Sky Star Map created by Annette Lee and collaborators as part of the *Native Skywatchers Middle School Teachers Workshop* held June 2012.

## 1.   Introduction

In Ojibwe the Big Dipper is *Ojiig* – the Fischer (Morton & Gawboy 2000; Gawboy 2005) and in D(L)akota star knowledge the same group of stars is seen as *To Win/Ton Win* – Blue Woman/Birth Woman (Goodman 1992). In each there are stories and teachings that help guide and inspire native peoples. The *Native Skywatchers Project* focuses on understanding the Ojibwe and D(L)akota importance of this and other celestial connections.

There is urgency to this project for two reasons: the native star knowledge is disappearing as elders pass and state standards. MN State Science Standards K-12 requires "Understanding that men and women throughout the history of all cultures, including Minnesota American Indian tribes and communities, have been involved in engineering design and scientific inquiry….For example Ojibwe and Dakota knowledge and use of patterns in the stars to predict and plan. " (Minnesota Department of Education, 2010) And yet there is a complete void of materials.

This research, *The Native Skywatchers Project,* seeks out elders and culture teachers to discuss the Ojibwe & D(L)akota star knowledge. From these sources and working with the elders we have created an astronomically accurate, culturally important Ojibwe star map, *Ojibwe Giizhig Anung Masinaaigan*. This valuable map, was disseminated to regional educators at the *Native Skywatchers Middle School Teacher* workshop June 2012. In addition, hands-on curriculum that combines astronomy, culture, language and art has been developed. The goal of the *Native Skywatchers* programming is to build community around the native star knowledge.

## 2.   Methods

### 2.1  Funding and Motivation

The *Native Skywatchers* programming is funded by: NASA-MN Space Grant, North Star STEM Alliance, St. Cloud State University and Fond du Lac Tribal & Community College. This project represents a unique collaboration between a large state university, a tribal and community college and a federal agency. The strategy of the project was to combine astronomical expertise, cultural knowledge and artistic talents to create an Ojibwe star map that currently does not exist. Regional teachers requesting information relating to the native star knowledge also motivation the map and related curriculum. This was clearly an outcome of the MN State Science Standards K-12, in particular benchmark 3.1.3.2.1 "Understand that everybody can use evidence to learn about

the natural world, identify patterns in nature, and develop tools." (Minnesota Department of Education, 2010)

### 2.2 Collaborators

Travel and interviews were initially conducted to consult with various Ojibwe cultural experts such as Carl Gawboy (Boise Fort), Paul Schultz (White Earth), and William Wilson (Lake Nipigon). The finished map is a collaborative work between Annette Lee, William Wilson and Carl Gawboy.

Annette Lee is an astrophysict and an artist. Currently an assistant professor of Physics and Astronomy and the Planetarium Director at St. Cloud State University. Annette is Dakota-Sioux and is a former faculty member at Fond du Lac Tribal & Community College. The *Native Skywatchers* programming is the focus of her research which combines twenty years of education and community involvement with both D(L)akota and Ojibwe communities. Recently awarded a Bush Fellowship to continue and expand the *Native Skywatchers* programming.

William Wilson is Ojibwe from Jack Fish Pine Reservation near Lake Nipigon, Ontario. William grew up traditionally, was raised by his grandparents speaking Ojibwe only. Since moving to the United States over eighteen years ago William has participated in various Ojibwe language revitalization projects such as *Waadookodaading*- Ojibwe Language Immersion School in Hayward Wisconsin, as an Ojibwe language expert. More recently William received recognition for his Ojibwe x-ray style paintings. His paintings have been shown regionally including a solo show at the Tweed Museum in Duluth where many paintings were purchased for the permanent collection. In addition, he recently received a MN State Arts Board-Cultural Community Partnership Grant to further develop his artistic talents.

Carl Gawboy is a retired professor of American Indian Studies from College of St. Scholastica in Duluth, MN and a distinguished professional artist. His work includes many traditional scenes of Ojibwe lifestyle purchased by the Tweed Museum for their permanent collection and murals of Ojibwe legends at various locations state wide. A member of the Boise-Fort Band of Ojibwe, Carl has also written books and plays including an international play examining the harsh realities of boarding school, *The Great Hurt*. Lorraine Norrgard has created a video titled *Carl Gawboy Portrait: The Art of Everyday*. Recently Carl won first prize in the Northeast Minnesota Book Awards for 2011 for his book *Ancient Earth and the First Ancestors*.

Carl Gawboy grew up in Ely Minnesota where bass fishing in Lake Hegman, Boundary Waters Canoe Area, Minnesota was a common excursion. Many people have visited the various pictographs throughout the region, but an understanding of the paintings remained a mystery. After decades of years of research (both formal and informal) on the Ojibwe star knowledge, Carl was able to connect traditional Ojibwe stories and teachings with some of the pictographs. His book, *Talking Rocks,* (Morton & Gawboy 2000) cowritten with geologist Ron Morton, explains in detail the connections between the Lake Hegman pictographs and Ojibwe constellations. More specifically, these particular rock paintings seen in Figure 1 are connected to three important Ojibwe legendary and seasonal figures: Curly Tail, Wintermaker, and Moose. There are hundreds of pictographs in this region, many of which are not fully understood (Furtman 2000).

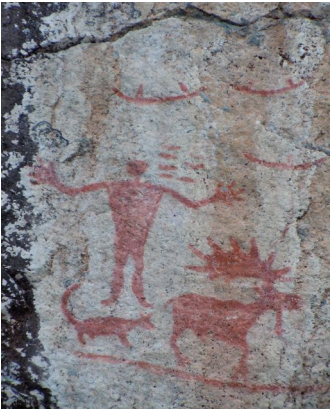

Figure 1. Pictographs at Lake Hegman.

### 2.3 Map Design

The original painting for the star map was created on Rives BFK 100% cotton, hot press 140lb paper 36" x 36". Completed as a mixed media piece, it combines drawing, watercolor, marker, colored pencil and pen & ink. Work began as an astronomical sketch of the celestial sphere and grid. Major celestial coordinates and stars of greater apparent magnitude were developed. William Wilson painted the Ojibwe constellations according to star chart created by Carl Gawboy. Annette Lee did all remaining artwork for the star map painting, including paintings of Greek constellations, borders and the seasonal layout.

After the Ojibwe constellations were set in place the Greek constellations were painted in a transparent naples yellow hue. An important design decision was made to allow the Ojibwe constellations to be on the top layer and the Greek constellations illustrated as if in a background layer. This is to recognize that many people are familiar with at least some of the Greek constellations and it is important to use this as a reference point.

Finally the Ojibwe star vocabulary was added to the mixed media Ojibwe star painting. William Wilson, an Ojibwe first language speaker and cultural expert, is the Ojibwe language consultant for this project. The star vocabulary is arranged according to the four seasons in the four corners of the map.

### 3. Results

#### 3.1 *Ojibwe Giizhig Anung Masinaaigan* – Ojibwe Sky Star Map

The map is organized with Polaris –the North Star in the center. This is to emphasize the closeness of Polaris to our current north celestial pole (NCP) and circumpolar motion. Because of circumpolar motion, we appear to see all the stars in the night sky revolve around the North Star in a counter-clock wise motion as the hours pass each night into day. Because of this motion, in some native cultures the North Star is seen as one of the leaders of the star nation. Stars near the North Star do not set below the horizon. These are referred to as North/Circumpolar Stars.

All other stars will rise in the east and set in the west at regular times throughout the year. They are seasonal stars. The *Ojibwe Giizhig Anung Masinaaigan* – Ojibwe Sky Star Map is arranged in order to show the constellations that are most visible each season. This assumes a viewing time of about three hours after sunset. In the actual night sky these stars would be seen overhead or in the south. For example if you look at the stars in the early summertime, a few hours after sunset you will see Hercules overhead and Scorpio low on the southern horizon. These are early summer stars.

The Ojibwe x-ray style used by William Wilson is symbolic of seeing the unseen. It is an allegory for the material world and the spirit world. The brightly colored internal organs and anatomical shapes are a glimpse into the inner mysteries of a being. "We are seeing the picture as the spirits see us. They see right through. The strange looking animals and figures are portrayed as they come in ceremony. Sometimes they are half beaver, half eagle. Sometimes they are scary. Sometimes tempting. ", explains William Wilson.

The border includes: strawberries, raspberries, blueberries and winterberries illustrated in reference to traditional Ojibwe style floral beadwork. Often the floral beadwork is done on black velvet or with a white beadwork background. Usually beadwork is done on items of importance spiritually or socially like pipe bags, moccasins, leggings, etc.

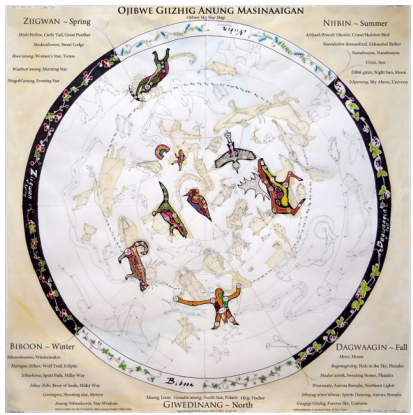

Figure 2. *Ojibwe Giizhig Anung Masinaaigan* – Ojibwe Sky Star Map created by Annette S. Lee, William Wilson, Carl Gawboy

### 3.2 Ojibwe Constellation Guide

Accompanying this map is the "Ojibwe Constellation Guide", written by Annette Lee, William Wilson, Carl Gawboy and Jeff Tibbetts. The guide is arranged in order of the seasons. The northern constellations are circumpolar and therefore do not set below the horizon (as seen from mid-northern latitudes) and are listed separately following the seasons.

*Ziigwan Anung* – **Spring Stars**

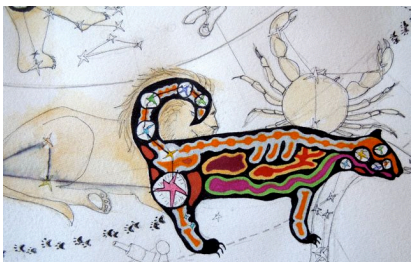

Figure 3. Close up of *Mishi Bizhiw* – Curly Tail constellation

*Mishi Bizhiw* - Curly Tail, Great Panther - Leo, Cancer.
This constellation is a mountain lion/cougar/puma that was once more abundant in Minnesota. The big spirit cat lives at the bottom of lakes and can cause flooding or water danger. Curly Tail rises in late winter and is overhead in spring. People knew that when the great cat was overhead the ice would be thawing and it would be dangerous to walk across partially frozen lakes or rivers. People also knew that it was time to move from winter camp to sugar bush camp. At sugar bush, feasts and prayers are offered for the water spirits (including Curly Tail) and to all those relatives that did not survive the winter.

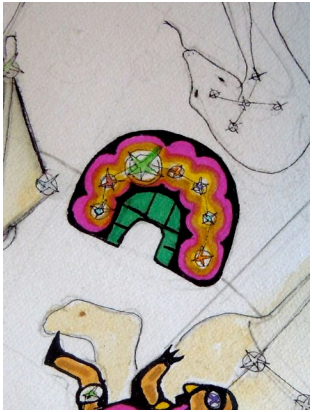

Figure 4. Close up of *Madoodiswan* – Sweat Lodge constellation

*Madoodiswan* - Sweat Lodge - Corona
The sweat lodge is a purification ceremony. It is returning to the womb and remembering/renewing of a person's spirit. The Sweat Lodge is seen overhead in late Spring.

**Niibin Anung – Summer Stars**

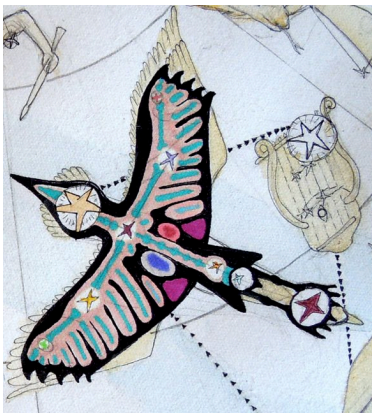

Figure 5. Close up of *Ajiijaak/Bineshi Okanin* – Crane/Skeleton Bird constellation

*Ajiijaak/Bineshi Okanin* - Crane/Skeleton Bird - Cygnus
The crane is one of the leaders in the Ojibwe clan system. Crane and loon lead the people to stay strong (Benton-Banai, 2010). This constellation is overhead a few hours after sunset in the summertime. Cranes can grow almost as tall as a person, with wingspans longer than most people. It is one of the tallest birds in the world and can fly very high.

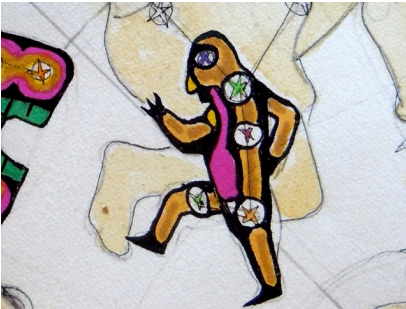

Figure 6. Close up of *Noondeshin Bemaadizid* – Exhausted Bather constellation

*Noondeshin Bemaadizid* - Exhausted Bather (Person) - Hercules

The sweat lodge is a purification ceremony. The person is physically exhausted after participating in the sweat, but full of life and renewed on the inside. The Exhausted Bather is an early summer constellation.

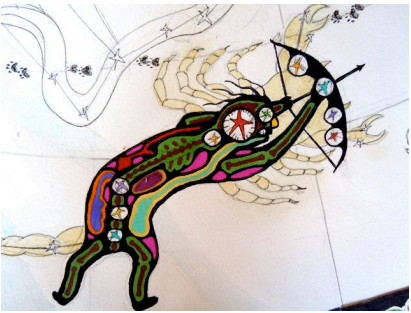

Figure 7. Close up of *Nanaboujou* – Nanaboujou constellation

*Nanaboujou* - Nanaboujou - Scorpio

A hero figure and a spirit that had many excursions on Earth a long time ago. He helped the people by creating dry land after the last flood. He had many human characteristics, like making mistakes. This constellation shows Nanaboujou shooting an arrow at the Great Panther, Curly Tail. There are many important Nanaboujou stories that are traditionally told only when there is snow on the ground.

*Dagwaagin Anung* – **Fall Stars**

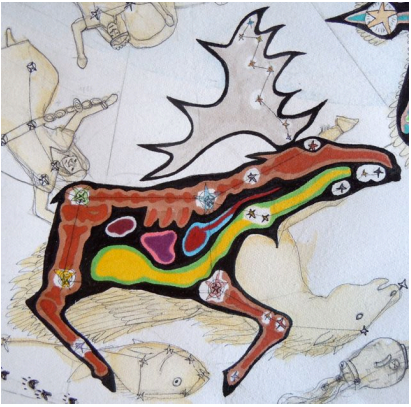
Figure 8.  Close up of *Mooz* – Moose constellation

*Mooz* - Moose - Pegasus, Lacerta
  This constellation is an animal of the Ojibwe clan system.  The moose provides food, clothing, and shelter for the people, much like deer or caribou. Compared to a deer, moose are much bigger and spend a lot of time in the water, especially in summer.  Male moose have a 'beard' of skin under their chin that can be seen in the constellation.

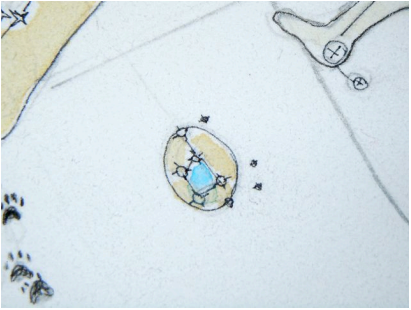
Figure 9.  Close up of *Bugonagiizhig* – Hole in the Sky constellation

*Bugonagiizhig* – Hole in the Sky & *Madoo'asinik* – Sweating Stones – Pleiades
  The hole in the sky refers to the Ojibwe shaking tent ceremony.  The tent that the medicine person builds acts as a spiritual doorway that relates to the Pleiades open star cluster.  In the sweat lodge purification ceremony basaltic rocks are heated in a central fire and then brought into the center of the lodge.  The Pleiades is seen overhead in late Fall.

  ***Biboon Anung* – Winter Stars**

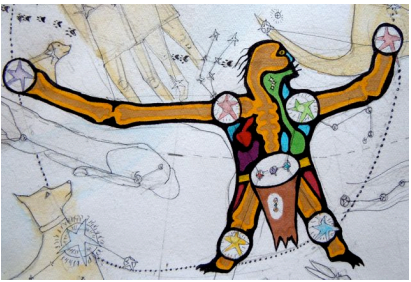
Figure 10.  Close up of *Biboonkeonini* – Wintermaker constellation

*Biboonkeonini* - Wintermaker - Orion, Procyon (Canis Minor), Aldebaran (Taurus)

Wintermaker is a strong Ojibwe canoe man (Gawboy 1992: Morton & Gawboy 2000: Gawboy 2005). His outstretched arms rule the winter sky. Wintermaker is seen overhead during the winter months. He is an important mythological figure in Ojibwe culture.

*Giwedinang Anung* – **North/Circumpolar Stars**

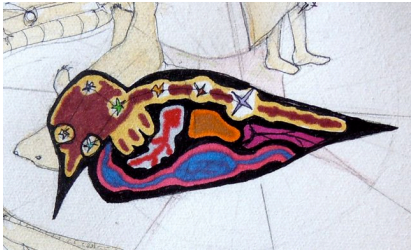

Figure 11. Close up of *Maang* – Loon constellation

Maang - Loon - Little Dipper/Ursa Minor

The North Star/Polaris and other stars in the Little Dipper/Ursa Minor make up the Ojibwe loon constellation. The loon is one of the Ojibwe clans and is seen as an important messenger and leader. The loon stands at the doorway between the water and the land but also between the material world and the spirit world. The loons' legs and feet are positioned very far back and they can't walk well on land therefore they avoid leaving the water. Preferring larger lakes, the loon only goes on land to nest.

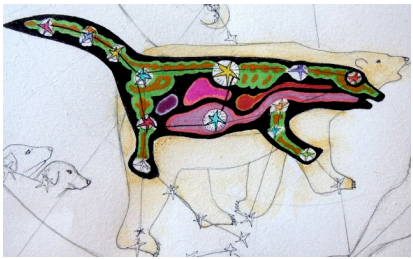

Figure 12. Close up of *Ojiig* – Fischer constellation

Ojiig - Fischer - Big Dipper/Ursa Major

This constellation relates to the story of when the birds and spring were held prisoner by the ogres. Of all the animals it was only the Fischer that was able to trick the ogres and free the birds. He saved everyone with his courage and wit. The fischer is the only animal that can kill and eat porcupines. Also it is not diurnal or nocturnal, but prefers to always be on the move, sleeping and eating night and day. It does not build a home in one place and return as most animals do, but rather makes its den in different places.

### 3.3 Ojibwe Star Vocabulary

Presented here is the Ojibwe Star Vocabulary that accompanies *Ojibwe Giizhig Anung Masinaaigan* – Ojibwe Sky Star Map. William Wilson, an Ojibwe first language speaker and cultural expert, lead the compilation of this section. For consistency two Ojibwe dictionaries were relied on: Baraga (1992) and Nichols & Nyholm (1995).

Table 1. Ojibwe Constellations

| | Ojibwe | Related Greek Constellations |
|---|---|---|
| Fall | *Dagwaagin* | |
| Moose | *Mooz* | Pegasus |
| Hole in the Sky | *Bugonagiizhig* | Pleiades |
| Sweating Stones | *Madoo'asinik* | Pleiades |
| Winter | *Biboon* | |
| Wintermaker | *Biboonkeonini* | Orion, Canis Minor, Taurus |
| Spring | *Ziigwan* | |
| Curly Tail, Great Panther | *Mishi bizhiw* | Leo, Hydra |
| Sweat Lodge | *Madoodiswan* | Corona |
| Summer | *Niibin* | |
| Crane/Skeleton Bird | *Ajiijaak/Bineshi Okanin* | Cygnus |
| Exhausted Bather (Person) | *Noondeshin Bemaadizid* | Hercules |
| Nanaboujou | *Nanaboujou* | Scorpio |
| North | *Giwedinang* | |
| Loon | *Maang* | Little Dipper |
| Fischer | *Ojiig* | Big Dipper |
| North Star | *Giwedin'anung* | Polaris |

The following table, Table 2, presents selected Ojibwe vocabulary related to celestial objects (Price 2002). There are stories and additional teachings that are contained in many of these words that will be presented in a subsequent dissemination.

Table 2.  Ojibwe Star Vocabulary

| Celestial Object | Ojibwe |
|---|---|
| Star | *Anung* |
| Star World | *Anung aki* |
| Moon | *Dibik-giizis (Night Sun)* |
| Sun | *Giizis* |
| Sky | *Giizhig* |
| Venus | *Ikwe'anung (Women's Star)* |
| Venus - Evening Star | *Ningobi'anung* |
| Venus - Morning Star | *Waabun'anung* |
| Ecliptic | *Maingan Mikan (Wolf Trail)* |
| Milky Way | *Jiibaykona (Spirit Path)* |
| Milky Way | *Jiibay Ziibi (River of Souls)* |
| Meteor/shooting star | *Gwiingwa* |
| Universe | *Gaagige Giizhig (Forever Sky) and Ishpeming (the Sky Above)* |
| Aurora Borealis (Northern Lights) | *Waawaate* |
| Star Knowledge (Wisdom) | *Anung Nibwakawin* |
| (Sky) Star map | *Giizhig Anung Masinaaigan* |
| East | *Waabun* |
| West | *Ningobi* |
| North | *Giwedinang* |
| South | *Jawanong* |

## 4. Conclusions and Future Directions

The methods presented here are interdisciplinary. Astronomy, culture, art, language, are all represented. And yet the delivery of such an in-depth, interdisciplinary topic like indigenous astronomy can be overwhelming to students, adults or youth, that have grown up with light pollution, tall buildings, and computers. Unlike traditional native people, current members of society spend a lot of time indoors. Most people have some familiarity with the Big Dipper, Sun and Moon.

The delivery of this culturally rich material must be simple and yet allow for complexity and abstraction. To achieve this goal we first use the cultural framework of the four directions. The current night sky is subdivided into: north, east, south/overhead, and west. From the beginning of the discussion the cultural context is intact. The four directions are considered important framework and guideposts in native culture. This instructional approach builds on a sense of place that often native peoples are aware of (Semken 2005) and allows participants to connect the current night sky to sense of place. This technique grounds the complexity of the current night sky in the tangible and the simple, and yet allows for a multi-layered, circular learning approach. Following this approach allows for the widest range of participants to take part in the learning experience.

Furthermore, the stars and constellations can be best understood in terms of the four seasons. The discussion is simplified again by fixing the time as a few hours after sunset. This is the observing time, and is referred to as 'prime time' for stargazing. Only in the northern direction (assuming mid-northern latitudes) will the constellations remain above the horizon through the night (and day). When an observer faces the south direction he/she will see the current season of stars. The previous season will be seen setting in the west and the following season will be seen rising in the east. The *Ojibwe Giizhig Anung Masinaaigan* – Ojibwe Sky Star Map is best presented by transforming the discussion into an experiential, hands-on event. In addition, this highly visual, holistic and cooperative learning environment is more consistent with a traditional native learning style (Cleary & Peacock 1997).

Lastly, the *Native Skywatchers Project* is a collaborative approach. Users of these materials are urged to seek out elders and native community members to bring into the classroom. Materials represented here should be viewed as a beginning.

This map and related materials were presented at the *Native Skywatchers Middle School Teachers Workshop* June 2012 held at St. Cloud State University, St. Cloud, MN and Fond duc Lac Tribal and Community College, Cloquet, MN. The *Native Skywatchers* workshop (June 2012) and the *Native Skywatchers* course (Spring 2012) at St. Cloud State University were funded by NASA-MN Space Grant and St. Cloud State University in collaboration with Fond du Lac Tribal and Community College.

This work will be presented to the larger community at the Astronomical Society of the Pacific (ASP) Aug 2012 Tucson, AZ meeting, "Communicating Science: A national conference on science education & public outreach". Other related *Native Skywatchers* maps and curriculum materials will be presented at Science in Society Conference, Berkeley, CA Nov 2012. Meetings are currently scheduled with the Minnesota Department of Education in order to disseminate the *Ojibwe Giizhig Anung Masinaaigan* – Ojibwe Sky Star Map and related curriculum materials to educators statewide.

Visit the *Native Skywatchers* website for upcoming events and downloads: http://web.stcloudstate.edu/planetarium/native_skywatchers.html

**Acknowledgements**. *Miigwetch* to Carl Gawboy for working on the Ojibwe star knowledge research for forty years and graciously sharing his knowledge as part of this project. The *Native Skywatchers* project acknowledges the elders and others that have kept this star knowledge alive. Further acknowledgement to Paul Schultz (White Earth) who was also a collaborator with this project and passed away suddenly in 2011.